# SELF-EQUIVALENT VOTING RULES

HÉCTOR HERMIDA-RIVERA[1] • h.hermidarivera@hse.ru November 12, 2026

*In this paper, I introduce a novel stability axiom for stochastic voting rules, called self-equivalence, by which a voting rule must be able to induce—whenever applied to a suitable finite set of voting rules—a convex combination of voting rules that coincides with itself. I then show that the unique anonymous, efficient, neutral, and self-equivalent voting rule assigns to every voter equal probability of being a dictator (i.e., uniform random dictatorship). Since randomly choosing a dictator might be considered undesirable, this paper suggests one reason for democratic societies not to decide whether to replace their voting rule by applying it to arbitrary finite sets of voting rules if they also desire stability.*

KEYWORDS: *voting, uniform random dictatorship, self-equivalence* JEL CODES: *D71, D72*

## 1. INTRODUCTION

**Overview**. Every society faces the problem of selecting a voting rule with which to make collective choices. On the one hand, society's voters may want to have a voting rule that will survive the test of time, as frequent changes are disruptive and costly. On the other hand, they may be indifferent to change, in which case their voting rule might naturally evolve until becoming stable. Hence, there exist both normative and positive reasons to identify stable voting rules.

In this paper, I provide a new stability axiom for *stochastic voting rules*,[2] called self-equivalence, by which a neutral voting rule must be able to induce—whenever applied to a suitable finite set of voting rules—a convex combination of voting rules that coincides with itself. I then use this axiom to provide a novel axiomatic characterization of the uniform random dictatorship, which is the stochastic voting rule that assigns to every voter equal probability of being a dictator. The gist of this characterization is the addition of self-equivalence to three standard democratic axioms: anonymity, efficiency, and neutrality. According to these three axioms, voters' identities are irrelevant (anonymity), strictly Pareto dominated alternatives get zero probability (efficiency[3]), and the labels of the alternatives are irrelevant (neutrality).

Theorem 1 states that the unique anonymous, efficient, neutral, and self-equivalent voting rule is the uniform random dictatorship, while Remark 1 addresses the logical independence of the axioms. Since randomly choosing a dictator might be considered undesirable, Theorem 1 suggests one reason for democratic societies not to decide whether to replace their voting rule by applying it to arbitrary finite sets of voting rules if they also desire stability.

**Self-equivalence**. To formalize the idea of a society deciding whether to replace its voting rule, voters must be able to linearly order the voting rules available to them. Given any linear order profile over alternatives, each voter's linear order and utility function naturally induce a weak order on any finite set of voting rules in which said rules are ranked according to the von Neumann & Morgenstern (1944) expected utility of the lotteries they select. However, a weak order constructed this way is not linear whenever at least two distinct voting rules choose lotteries with identical expected utility. Then, a linear order is said to be *compatible with a weak order* if and only if it can be obtained from the weak order by linearly ordering the voting rules within each indifference class.

Now, a neutral voting rule is self-equivalent if and only if there exists a finite lottery over voting rules inducing a convex combination identical to the voting rule itself such that for every finite set of voting rules containing its support, every linear order profile over alternatives, and every utility profile representing it, there exists some compatible linear order profile at which the voting rule recovers said lottery.[4] Intuitively, if the support of the inducing lottery is available and ties are resolved so as to recover it, a self-equivalent voting rule reproduces itself when used to choose a lottery on voting rules. Thus, under these conditions, a society considering whether to replace a self-equivalent voting rule using itself will not do so. Therefore, self-equivalence can be seen as an *ex-ante* stability notion under which society adopts the voting rule induced by the recovered lottery instead of randomly choosing one of its components.

[1] *Faculty of Economic Sciences, HSE University, 11 Pokrovsky Bulvar, Moscow, 109028, Russia.*

[2] A *stochastic voting rule* is a function that assigns a lottery over alternatives to every linear order profile over alternatives.

[3] This property has also been called *ex-post Pareto efficiency*.

**Motivation**. There are both empirical and theoretical reasons to study stability in stochastic environments. From an empirical perspective, many voting rules used in practice are stochastic—even if only because of the way they resolve some ties. In fact, all neutral and anonymous voting rules choose some non-degenerate lottery at some linear order profile on some alternative set, for they cannot favor particular alternatives or voters. Given that neutrality and anonymity are staples of democratic systems, the stability of such systems cannot be fully addressed without considering stochasticity.

From a theoretical perspective, the resolute case was already analyzed by Koray (2000), who proposed a consistency property—called *universal self-selectivity*—by which a neutral voting rule must always be able to choose itself among any finite set of neutral voting rules it belongs to. Koray (2000, Theorem 2, p. 990) showed that a neutral and unanimous voting rule is universally self-selective if and only if it is dictatorial. Hence, one may wonder if a novel stability notion for stochastic voting rules might yield more democratic results than universal self-selectivity.

**Literature**. Koray's (2000) pioneering analysis has been extended by several authors. Some of them, such as Kerman & Koray (2023), Koray & Slinko (2008), and Koray & Ünel (2003), imposed different restrictions on the set of voting rules society can choose from. Other authors have developed this literature in different directions. Notably, Hermida-Rivera & Kerman (2025) introduced a binary variant of self-selectivity; Koray & Senocak (2024) proposed the notion of selection-closedness; Diss (2015) studied the manipulability of voting rules under self-selectivity conditions; Diss, Louichi, Merlin, & Smaoui (2012) and Diss & Merlin (2010) examined the stability of the triplet comprising plurality, Borda, and antiplurality; Lainé, Ozkes, & Sanver (2016) extended the notion of self-selectivity to social welfare functions; and Bhattacharya (2019) introduced the closely related notion of constitutional consistency.

Several characterizations of random dictatorships have been obtained.[5] Gibbard (1977, Corollary 1, p. 677) proved that with at least three alternatives, a voting rule is efficient and strategyproof if and only if it is a random dictatorship.[6] Subsequently, Gibbard (1978) considered arbitrary finite mechanisms and obtained, in Corollary 1 (1978, p. 612), a related random dictatorship result based on straightforwardness and attainability. Further, Pattanaik & Peleg (1986, Theorem 4.11, p. 917) obtained a characterization of random dictatorships on proper subsets of a universal set with at least four alternatives via efficiency, regularity, and a probabilistic counterpart of Arrow's (1950, 1951) independence of irrelevant alternatives. In a framework embedding anonymity and topsonliness, Ozkes & Sanver (2023, Theorem 1, p. 3) proved that with at least three alternatives, the uniform random dictatorship is characterized by efficiency and independence. Finally, Bandhu, Lahiri, & Pramanik (2024, Theorem 2(b), p. 177) showed that with at least four alternatives, random dictatorships are identified by efficiency, tops-onliness, and stochastic same-sidedness.

[4]For the self-equivalence axiom to be well-defined, a voting rule must be able to deal with nonempty alternative sets of any finite cardinality. Hence, this paper mirrors Koray (2000) by defining a voting rule as a function from the set of all linear order profiles on all finite nonempty alternative sets to the set of all lotteries over such alternative sets.

[5]Adding anonymity to any characterization of random dictatorships provides a characterization of the uniform random dictatorship, as anonymity forces every voter to be a dictator with equal probability.

[6]In doing so, he extended to stochastic voting rules the so-called Gibbard-Satterthwaite Theorem, by which in environments with at least three alternatives, a resolute voting rule is unanimous and strategyproof if and only if it is dictatorial (Gibbard, 1973, Corollary, p. 595; Satterthwaite, 1975, Theorem 1, p. 193).

**Notation**. Throughout the paper, I adopt the following notational conventions: (**1**) for any collection of individual objects $(\vartheta_i)_{i \in N}$, the corresponding profile is denoted by the unsubscripted letter $\vartheta = (\vartheta_i)_{i \in N}$; (**2**) unless otherwise specified, accented symbols denote objects of the same kind as their unaccented counterparts; (**3**) variable ranges in quantifiers and set-builder notation are omitted when clear from the context.

**Structure**. Section 2 sets out the environment, Section 3 defines the uniform random dictatorship, Section 4 introduces the base axioms, Section 5 formalizes self-equivalence, Section 6 provides some examples, Section 7 states and proves the results, and Section 8 discusses some key modeling choices.

## 2. Environment

The *environment* is a 6-tuple $(N, A_\tau, \mathcal{P}(A_\tau), \Delta(A_\tau), v, f)$. Let $N = \{1, \dots, n\}$ be a *finite set of voters* $i, j, k, \dots$, where $n \geqslant 2$. Given any $\tau \in \mathbb{N} = \{1, 2, \dots\}$, let $A_\tau = \{1, \dots, \tau\}$ be a *finite set of alternatives* $x, y, z, \dots$, let

$$\mathcal{P}(A_\tau) = \big\{(P_i)_{i \in N} \,\big|\, (\forall i)[P_i \text{ is a linear order on } A_\tau]\big\} \tag{1}$$

be the *set of all linear order profiles* $P$ *on the alternative set*,[7] and let

$$\Delta(A_\tau) = \big\{\delta : A_\tau \to [0, 1] \,\big|\, \textstyle\sum_{x \in A_\tau} \delta(x) = 1\big\} \tag{2}$$

be the *set of all lotteries* $\delta$ *on the alternative set*. Given any $\tau$ and $P$, let $v : A_\tau \to \mathbb{R}^N$ be any *utility function profile* satisfying, for all $i$, $x$, and $y$, $v_i(x) \geqslant v_i(y)$ if and only if $x P_i y$. Finally, let

$$f : \textstyle\bigcup_{\tau \in \mathbb{N}} \mathcal{P}(A_\tau) \to \bigcup_{\tau \in \mathbb{N}} \Delta(A_\tau) \tag{3}$$

be any *voting rule* such that $f(P) \in \Delta(A_\tau)$ for all $\tau$ and $P$.

Let $\mathcal{F} = \{f \mid f \text{ is a voting rule}\}$.

## 3. Uniform random dictatorship

I now introduce the voting rule to be characterized: the uniform random dictatorship. Given any $\tau$, $P$, $i$, and $m \in \{1, \dots, \tau\}$, let $t_m(P_i)$ be the *alternative in the* $m^{th}$ *position at* $P_i$. Given any $i$, let $i$'s *dictatorship* $d_i \in \mathcal{F}$ satisfy

$$(\forall \tau, P)[d_i(P)(t_1(P_i)) = 1] \tag{4}$$

Let $\mathcal{F}_d = \{d_i \mid i \in N\}$.

Definition 1 — Uniform random dictatorship. The *uniform random dictatorship* $\varphi$ assigns to every voter equal probability of being a dictator. Formally,

$$\varphi = (1/n) \textstyle\sum_{i \in N} d_i \tag{5}$$

[7] A *linear order* is a *complete*, *transitive*, and *antisymmetric* binary relation. Thus, unlike a *strict linear order*, it is also *reflexive*.

Given any $\tau$, $P$, $x$, and $m \in \{1, \dots, \tau\}$, let $N_m(x, P) = \{i \mid t_m(P_i) = x\}$ be the *set of all voters who rank $x$ in the $m^{th}$ position at* $P$, and let $n_m(x, P) = |N_m(x, P)|$ be *its cardinality*. Hence,

$$(\forall \tau, P, x)[\varphi(P)(x) = (1/n)n_1(x, P)] \quad 6$$

## 4. AXIOMS

I now introduce the anonymity, efficiency, and neutrality axioms, which are just natural adaptations to stochastic environments of their well-known resolute counterparts.

Let $\pi : N \longrightarrow N$ be any *permutation of the voter set*. Given any $\tau$, $P$, and $\pi$, let the *linear order profile* $\pi P$ satisfy

$$(\forall i, x, y)[(x(\pi P)_i y) \Leftrightarrow (x P_{\pi(i)} y)] \quad 7$$

AXIOM 1 — ANONYMITY. $f$ is *anonymous* if and only if it is independent of voters' identities. Formally,

$$(\forall \tau, P, \pi)[f(P) = f(\pi P)] \quad 8$$

Let $\mathcal{F}_a = \{f \mid f \text{ is anonymous}\}$.

AXIOM 2 — EFFICIENCY. $f$ is *efficient* if and only if it assigns zero probability to every strictly Pareto dominated alternative. Formally,

$$(\forall \tau, P, x, y)[((x \neq y) \wedge ((\forall i)[x P_i y])) \Rightarrow (f(P)(y) = 0)] \quad 9$$

Let $\mathcal{F}_e = \{f \mid f \text{ is efficient}\}$.

Given any $\tau$, let $\mu : A_\tau \longrightarrow A_\tau$ be any *permutation of the alternative set*. Given any $\tau$, $P$, and $\mu$, let the *linear order profile* $\mu P$ satisfy

$$(\forall i, x, y)[(\mu(x)(\mu P)_i \mu(y)) \Leftrightarrow (x P_i y)] \quad 10$$

AXIOM 3 — NEUTRALITY. $f$ is *neutral* if and only if it is independent of the labels of the alternatives. Formally,

$$(\forall \tau, P, \mu, x)[f(P)(x) = f(\mu P)(\mu(x))] \quad 11$$

Let $\mathcal{F}_n = \{f \mid f \text{ is neutral}\}$.

The domain of any $f \in \mathcal{F}_n$ can be easily extended to linear order profiles on any finite nonempty set. Given any $\tau$, let $X$ be any *finite nonempty set of elements* $a, b, \dots$ with cardinality $\tau$, let $L$ be any *linear order profile on* $X$, and let $\beta : X \longrightarrow A_\tau$ be any *bijection*. Given any $L$ and $\beta$, let the *linear order profile* $\beta L$ satisfy

$$(\forall i, a, b)[(\beta(a)(\beta L)_i \beta(b)) \Leftrightarrow (a L_i b)] \quad 12$$

Then, given any $f \in \mathcal{F}_n$,

$$(\forall a, L, \beta, \tilde{\beta})[f(\beta L)(\beta(a)) = f(\tilde{\beta} L)(\tilde{\beta}(a))] \quad 13$$

Hence, given any $\beta$ and $f \in \mathcal{F}_n$, extend $f$ by setting

$$(\forall a, L)[f(L)(a) = f(\beta L)(\beta(a))] \quad 14$$

Similarly, given any $\beta$ and $m \in \{1, \dots, \tau\}$, set

$$(\forall i, L)[t_m(L_i) = \beta^{-1}(t_m((\beta L)_i))] \quad 15$$
$$(\forall a, L)[N_m(a, L) = N_m(\beta(a), \beta L)] \quad 16$$
$$(\forall a, L)[n_m(a, L) = n_m(\beta(a), \beta L)] \quad 17$$

## 5. SELF-EQUIVALENCE

In order to introduce the self-equivalence axiom, I first show how utility profiles on alternative sets induce weak order profiles on nonempty finite sets of voting rules through the expected utilities of the lotteries these select. Given any $\tau$, $P$, and $\nu$, let $u : \Delta(A_\tau) \to \mathbb{R}^N$ be the *expected utility function profile* that satisfies, for all $\delta$ and $i$,

$$u_i(\delta) = \sum_{x \in A_\tau} \delta(x)\nu_i(x) \quad 18$$

Let $\Gamma \subsetneq \mathcal{F}$ be any *finite nonempty set of voting rules* $g$. Given any $\tau$, $P$, $\nu$, and $\Gamma$, let the *weak order profile* $R^\nu$ satisfy

$$(\forall i, g, \tilde{g})[(g R_i^\nu \tilde{g}) \Leftrightarrow (u_i(g(P)) \geqslant u_i(\tilde{g}(P)))] \quad 19$$

and let $R$ be any *linear order profile on* $\Gamma$ satisfying $R_i \subseteq R_i^\nu$ for all $i$. Then, any such $R$ is a *linear order profile compatible with* $R^\nu$. Given any $f \in \mathcal{F}_n$, let $\lambda : \mathcal{F} \to [0, 1]$ be any *finite lottery* satisfying

$$f = \sum_{h \in \mathcal{F}} \lambda(h)h \quad 20$$

Then, any such $\lambda$ is a *finite lottery inducing* $f$. Given any such $\lambda$, let $G \subsetneq \mathcal{F}$ be any *finite set satisfying* $\text{supp}(\lambda) \subseteq G$.

AXIOM 4 — SELF-EQUIVALENCE. $f \in \mathcal{F}_n$ is *self-equivalent* if and only if there exists a finite lottery inducing $f$ such that, for every linear order profile, utility profile, and finite set of voting rules containing the support of that lottery, there exists a compatible linear order profile on that set at which $f$ recovers that lottery. Formally,

$$(\exists \lambda)(\forall \tau, P, \nu, G)(\exists R)[f(R) = \lambda|_G] \quad 21$$

Let $\mathcal{F}_s = \{f \in \mathcal{F}_n \mid f \text{ is self-equivalent}\}$.

## 6. EXAMPLES

Examples 1 to 4 illustrate the intuition underpinning the results of this paper. Let $N = \{1, \ldots, 4\}$, let $\tau = 3$, and let $P$ be as given in Table 1. Examples 1 to 3 use the following construction. Consider any $f \in \mathcal{F}_n$ satisfying $f(P)(2) = 1$, any $\nu$, and any $\lambda$ inducing $f$. Let $G = \text{supp}(\lambda) \cup \{d_1\}$. Since $\lambda$ induces $f$ and $f(P)(2) = 1$, it follows that $g(P)(2) = 1$ for all $g \in \text{supp}(\lambda)$. Further, $d_1(P)(1) = 1$. Hence, $d_1 \in G \setminus \text{supp}(\lambda)$ and $\lambda(d_1) = 0$. Then, $R^\nu$ is depicted in Table 2.

TABLE 1. LINEAR ORDER PROFILE $P$

| $P_1$ | $P_2$ | $P_3$ | $P_4$ |
|---|---|---|---|
| 1 | 2 | 3 | 2 |
| 2 | 3 | 1 | 3 |
| 3 | 1 | 2 | 1 |

TABLE 2. WEAK ORDER PROFILE $R^{\nu}$ FOR $f \in \{p, b, c\}$

| $R_1^{\nu}$ | $R_2^{\nu}$ | $R_3^{\nu}$ | $R_4^{\nu}$ |
|---|---|---|---|
| $d_1$ | $\text{supp}(\lambda)$ | $d_1$ | $\text{supp}(\lambda)$ |
| $\text{supp}(\lambda)$ | $d_1$ | $\text{supp}(\lambda)$ | $d_1$ |

EXAMPLE 1 — PLURALITY RULE. The *plurality rule* $p$, which assigns equal probability to all alternatives that are top-ranked by the largest number of voters and zero probability to all other alternatives, is not self-equivalent. Since $n_1(1, P) = 1$, $n_1(2, P) = 2$, and $n_1(3, P) = 1$, it follows that $p(P)(2) = 1$. Thus, consider any $R$. Then, $n_1(d_1, R) = 2$ and $n_1(g, R) \leqslant 2$ for all $g \in \text{supp}(\lambda)$. Hence, $p(R)(d_1) \in \{1/2, 1\}$, $p(R)(d_1) > 0 = \lambda(d_1)$, and $p(R) \neq \lambda|_G$. Therefore, $p$ is not self-equivalent.

EXAMPLE 2 — BORDA RULE. The *Borda rule* $b$ with scoring vector $(\tau, \tau - 1, \dots, 1)$, which assigns equal probability to all alternatives with the largest Borda score and zero probability to all other alternatives, is not self-equivalent. Given any $x$, let the *Borda score of* $x$ *at* $P$ be

$$s(x, P) = \textstyle\sum_{m=1}^{\tau}(\tau - m + 1)n_m(x, P) \tag{22}$$

Then, $s(1, P) = 7$, $s(2, P) = 9$, and $s(3, P) = 8$. Therefore, $b(P)(2) = 1$. Suppose there exists some $R$ satisfying $b(R)(g) = \lambda(g)$ for all $g$. Fix any such $R$, and let $\alpha = |G|$. Given any $g$, let the *Borda score of* $g$ *at* $R$ be

$$s(g, R) = \textstyle\sum_{m=1}^{\alpha}(\alpha - m + 1)n_m(g, R) \tag{23}$$

Since $n_1(d_1, R) = n_\alpha(d_1, R) = 2$, it follows that $s(d_1, R) = 2\alpha + 2$. Further, $b(R)(g) = \lambda(g) > 0$ for all $g \in \text{supp}(\lambda)$. Hence, $s(g, R) = s(\tilde{g}, R)$ for all $g, \tilde{g} \in \text{supp}(\lambda)$. Thus,

$$\textstyle\sum_{g \in G} s(g, R) = 4\sum_{m=1}^{\alpha} m = 2\alpha^2 + 2\alpha \tag{24}$$

Since $|\text{supp}(\lambda)| = \alpha - 1 \geqslant 1$ and $s(d_1, R) = 2\alpha + 2$, it follows that $s(g, R) = (2\alpha^2 - 2)/(\alpha - 1) = 2\alpha + 2$ for all $g$. Hence, $b(R)(d_1) = 1/\alpha > 0 = \lambda(d_1)$, which is a contradiction. Then, $b(R) \neq \lambda|_G$ for all $R$. Therefore, $b$ is not self-equivalent.

EXAMPLE 3 — WEAK CONDORCET RULE. The *weak Condorcet rule* $c$, which assigns equal probability to all alternatives that are ranked above each other alternative by at least half of the voters (if such alternatives exist) and equal probability to all strictly Pareto undominated alternatives otherwise, is not self-equivalent. Then,

$$|\{i \mid 2P_i1\}| = |\{i \mid 1P_i2\}| = 2 \tag{25}$$

$$|\{i \mid 2P_i3\}| = 3 > 1 = |\{i \mid 3P_i2\}| \tag{26}$$

$$|\{i \mid 3P_i1\}| = 3 > 1 = |\{i \mid 1P_i3\}| \tag{27}$$

Therefore, $c(P)(2) = 1$. Thus, consider any $R$. Then, $|\{i \mid d_1R_ig\}| = 2$ for all $g \in \text{supp}(\lambda)$. Hence, $c(R)(d_1) > 0 = \lambda(d_1)$ and $c(R) \neq \lambda|_G$. Therefore, $c$ is not self-equivalent.

EXAMPLE 4 — UNIFORM RANDOM DICTATORSHIP. The uniform random dictatorship $\varphi$, which assigns to every voter equal probability of being a dictator, satisfies the predicate of self-equivalence at $P$. Let $\lambda$ satisfy $\lambda(d_i) = 1/4$ for all $i$. Consider any $\nu$ and $G$. Then, $d_1(P)(1) = d_2(P)(2) = d_4(P)(2) = d_3(P)(3) = 1$. Given any $i$ and $g$, $u_i(g(P)) \leqslant \nu_i(t_1(P_i)) = u_i(d_i(P))$. Thus, $d_iR_i^{\nu}g$ for all $i$ and $g$. Consequently, there exists some $R$ satisfying $t_1(R_i) = d_i$ for all $i$. Fix any such $R$. Then, $n_1(d_i, R) = 1$ for all $i$, whereas $n_1(g, R) = 0$ for all $g \in G \setminus \mathcal{F}_d$. By equation 6, $\varphi(R)(g) = (1/4)n_1(g, R) = \lambda(g)$ for all $g$. Hence, $\varphi(R) = \lambda|_G$. Therefore, $\varphi$ satisfies the predicate of self-equivalence at $P$.

## 7. Results

THEOREM 1. *The unique anonymous, efficient, neutral, and self-equivalent voting rule is the uniform random dictatorship. Formally,*

$$\mathcal{F}_a \cap \mathcal{F}_e \cap \mathcal{F}_n \cap \mathcal{F}_s = \{\varphi\} \tag{28}$$

PROOF OF THEOREM 1. There are two statements to show:

1. $\{\varphi\} \subseteq \mathcal{F}_a \cap \mathcal{F}_e \cap \mathcal{F}_n \cap \mathcal{F}_s$,
2. $\mathcal{F}_a \cap \mathcal{F}_e \cap \mathcal{F}_n \cap \mathcal{F}_s \subseteq \{\varphi\}$.

STATEMENT 1. $\{\varphi\} \subseteq \mathcal{F}_a \cap \mathcal{F}_e \cap \mathcal{F}_n \cap \mathcal{F}_s$.

The proof of Statement 1 follows from Claims 1.1 to 1.4.

CLAIM 1.1. $\varphi \in \mathcal{F}_a$.

The proof is direct. Consider any $\tau$, $\pi$, $P$, and $x$. Then, $n_1(x, P) = n_1(x, \pi P)$. By equation 6, $\varphi(P)(x) = (1/n)n_1(x, P) = (1/n)n_1(x, \pi P) = \varphi(\pi P)(x)$. Hence, $\varphi \in \mathcal{F}_a$.

CLAIM 1.2. $\varphi \in \mathcal{F}_e$.

The proof is direct. Consider any $\tau$, $P$, $x$, and $y$ such that $x \neq y$ and $xP_iy$ for all $i$. Then, $N_1(y, P) = \emptyset$. By equation 6, $\varphi(P)(y) = (1/n)n_1(y, P) = (1/n)0 = 0$. Hence, $\varphi \in \mathcal{F}_e$.

CLAIM 1.3. $\varphi \in \mathcal{F}_n$.

The proof is direct. Consider any $\tau$, $P$, $x$, and $\mu$. Then, $N_1(x, P) = N_1(\mu(x), \mu P)$. By equation 6, $\varphi(P)(x) = (1/n)n_1(x, P) = (1/n)n_1(\mu(x), \mu P) = \varphi(\mu P)(\mu(x))$. Hence, $\varphi \in \mathcal{F}_n$.

CLAIM 1.4. $\varphi \in \mathcal{F}_s$.

The proof is direct. By equation 5, let $\lambda$ satisfy $\lambda(d_i) = 1/n$ for all $i$. Then, $\text{supp}(\lambda) = \mathcal{F}_d$. Consider any $\tau$, $P$, $\nu$, and $G$. Given any $i$ and $g$, $u_i(g(P)) \leqslant \nu_i(t_1(P_i)) = u_i(d_i(P))$. Hence, $d_i R_i^\nu g$ for all $i$ and $g$. Therefore, there exists some $R$ such that $t_1(R_i) = d_i$, $N_1(d_i, R) = \{i\}$, and $n_1(d_i, R) = 1$ for all $i$. Fix any such $R$. By equation 6, $\varphi(R)(d_i) = (1/n)n_1(d_i, R) = 1/n = \lambda(d_i)$ for all $i$. Thus, $\varphi(R) = \lambda|_G$. Hence, $\varphi \in \mathcal{F}_s$.

STATEMENT 2. $\mathcal{F}_a \cap \mathcal{F}_e \cap \mathcal{F}_n \cap \mathcal{F}_s \subseteq \{\varphi\}$.

Let $f \in \mathcal{F}_a \cap \mathcal{F}_e \cap \mathcal{F}_n \cap \mathcal{F}_s$. By self-equivalence, there exists some $\lambda$ such that

$$(\forall \tau, P, \nu, G)(\exists R)[f(R) = \lambda|_G] \tag{29}$$

Fix any such $\lambda$, let $H = \text{supp}(\lambda)$, and let $h$ be any *voting rule in* $H$. Given any $\tau$ and $P$, let $B(P) = \{t_1(P_i) \mid i \in N\}$ be the *set of all alternatives top-ranked by at least one voter*, and given any $G$ and $R$, let $B(R) = \{t_1(R_i) \mid i \in N\}$ be the *set of all voting rules top-ranked by at least one voter*. The proof of Statement 2 follows from Claims 2.1 to 2.5.

CLAIM 2.1. $(\forall \tau, P)[\text{supp}(f(P)) \subseteq B(P)]$.

The proof is by contradiction. Suppose there exist some $\tilde{\tau}$ and $\tilde{P}$ such that $\text{supp}(f(\tilde{P})) \not\subseteq B(\tilde{P})$. Fix any such $\tilde{\tau}$ and $\tilde{P}$. Since $\lambda$ induces $f$, there exists some $h$ such that

$$\textstyle\sum_{x \in A_{\tilde{\tau}} \setminus B(\tilde{P})} h(\tilde{P})(x) > 0 \tag{30}$$

Fix any such $h$. Now, let the function

$$\tilde{h} : \bigcup_{\tau\in\mathbb{N}} \mathcal{P}(A_\tau) \to \bigcup_{\tau\in\mathbb{N}} \Delta(A_\tau) \tag{31}$$

satisfy, for all $\tau$, $P$, and $x$,

$$\tilde{h}(P)(x) = \begin{cases} h(P)(x) + r(P)\varphi(P)(x) & \text{if } x \in B(P) \\ 0 & \text{if } x \in A_\tau \setminus B(P) \end{cases} \tag{32}$$

where

$$r(P) = \sum_{x\in A_\tau\setminus B(P)} h(P)(x) \tag{33}$$

Then, $\tilde{h}(P)(x) \geqslant 0$ for all $\tau$, $P$, and $x$. Given any $\tau$ and $P$, equation 34 follows from equation 32, equation 36 follows from equation 6, equation 37 follows from $B(P) \subseteq A_\tau$, and equation 38 follows from $h \in \mathcal{F}$ and equation 33:

$$\begin{aligned} \sum_{x\in A_\tau} \tilde{h}(P)(x) &= \sum_{x\in B(P)}(h(P)(x) + r(P)\varphi(P)(x)) && (34)\\ &= \sum_{x\in B(P)} h(P)(x) + r(P) \sum_{x\in B(P)} \varphi(P)(x) && (35)\\ &= \sum_{x\in B(P)} h(P)(x) + r(P) && (36)\\ &= \sum_{x\in A_\tau} h(P)(x) - \sum_{x\in A_\tau\setminus B(P)} h(P)(x) + r(P) && (37)\\ &= 1 - r(P) + r(P) && (38)\\ &= 1 && (39) \end{aligned}$$

Thus, $\tilde{h} \in \mathcal{F}$. Since $r(\tilde{P}) > 0$, it follows that $\tilde{\tau} \geqslant 2$. Let $v$ satisfy

$$(\forall i)[v_i(t_1(\tilde{P}_i)) = 1] \tag{40}$$

$$(\forall i)(\forall m \in \{2, \dots, \tilde{\tau}\})[v_i(t_m(\tilde{P}_i)) = (\tilde{\tau} - m)/(2n\tilde{\tau})] \tag{41}$$

Given any $i$ and $m \in \{2, \dots, \tilde{\tau}\}$, it follows that

$$0 \leqslant v_i(t_m(\tilde{P}_i)) = (\tilde{\tau} - m)/(2n\tilde{\tau}) \leqslant (\tilde{\tau} - 2)/(2n\tilde{\tau}) < \tilde{\tau}/(2n\tilde{\tau}) = 1/(2n) \tag{42}$$

Hence, $0 \leqslant v_i(x) < 1/(2n)$ for all $i$ and $x \neq t_1(\tilde{P}_i)$. Given any $i$, equation 43 follows from equation 40, equation 44 follows from equation 41, equation 45 follows from equation 6, and equation 46 follows from $n_1(t_1(\tilde{P}_i), \tilde{P}) \geqslant 1$:

$$\begin{aligned} u_i(\varphi(\tilde{P})) &= \sum_{x\in A_{\tilde{\tau}}\setminus\{t_1(\tilde{P}_i)\}} v_i(x)\varphi(\tilde{P})(x) + \varphi(\tilde{P})(t_1(\tilde{P}_i)) && (43)\\ &\geqslant \varphi(\tilde{P})(t_1(\tilde{P}_i)) && (44)\\ &= (1/n)n_1(t_1(\tilde{P}_i), \tilde{P}) && (45)\\ &\geqslant 1/n && (46) \end{aligned}$$

Now, equation 32 implies that

$$\tilde{h}(\tilde{P})(x) - h(\tilde{P})(x) = \begin{cases} r(\tilde{P})\varphi(\tilde{P})(x) & \text{if } x \in B(\tilde{P}) \\ -h(\tilde{P})(x) & \text{if } x \in A_{\tilde{\tau}} \setminus B(\tilde{P}) \end{cases} \tag{47}$$

Thus, given any $i$, equation 49 follows from $B(\tilde{P}) \subseteq A_{\tilde{\tau}}$ and equation 47; equation 50 follows from equation 6; equation 51 follows from equation 46, $r(\tilde{P}) > 0$, and the bounds $v_i(x) < 1/(2n)$ and $h(\tilde{P})(x) \geqslant 0$ for all $x \in A_{\tilde{\tau}} \setminus B(\tilde{P})$; equation 52 follows from equation 33; and equation 54 follows from $r(\tilde{P}) > 0$ and $n \geqslant 2$:

$$u_i(\tilde{h}(\tilde{P})) - u_i(h(\tilde{P})) = \sum_{x \in A_{\tilde{\tau}}} v_i(x)(\tilde{h}(\tilde{P})(x) - h(\tilde{P})(x)) \quad (48)$$
$$= r(\tilde{P}) \sum_{x \in B(\tilde{P})} v_i(x)\varphi(\tilde{P})(x) - \sum_{x \in A_{\tilde{\tau}} \setminus B(\tilde{P})} v_i(x)h(\tilde{P})(x) \quad (49)$$
$$= r(\tilde{P})u_i(\varphi(\tilde{P})) - \sum_{x \in A_{\tilde{\tau}} \setminus B(\tilde{P})} v_i(x)h(\tilde{P})(x) \quad (50)$$
$$\geqslant r(\tilde{P})/n - (1/(2n)) \sum_{x \in A_{\tilde{\tau}} \setminus B(\tilde{P})} h(\tilde{P})(x) \quad (51)$$
$$= r(\tilde{P})/n - r(\tilde{P})/(2n) \quad (52)$$
$$= r(\tilde{P})/(2n) \quad (53)$$
$$> 0 \quad (54)$$

Let $G = H \cup \{\tilde{h}\}$. By self-equivalence, there exists some $R$ satisfying $f(R)(h) = \lambda(h) > 0$. Fix any such $R$. By equation 54, $h \neq \tilde{h}$ and $\tilde{h} R_i h$ for all $i$. Then, efficiency implies that $f(R)(h) = 0$. Thus, $0 = f(R)(h) = \lambda(h) > 0$, which is a contradiction. Hence, $\text{supp}(f(P)) \subseteq B(P)$ for all $\tau$ and $P$.

CLAIM 2.2. $(\forall \tau, P, h)(\exists i)[h(P)(t_1(P_i)) = 1]$.

The proof is direct. Consider any $\tau$, $P$, and $v$. Let $G = H \cup \mathcal{F}_d$. By self-equivalence, there exists some $R$ such that $\text{supp}(f(R)) = H$. Fix any such $R$. Given any $h$, Claim 2.1 implies that $t_1(R_i) = h$ for some $i$. Fix any such $i$. Then, $h R_i^v d_i$. But since $d_i R_i^v h$ and $v_i(x) < v_i(t_1(P_i))$ for all $x \neq t_1(P_i)$, it follows that $h(P) = d_i(P)$. Hence, $h(P)(t_1(P_i)) = 1$.

CLAIM 2.3. $|H| = n$.

The proof is direct. Let $\tau = n$, and let $\mu$ and $\pi$ satisfy

$$((\forall x \in \{1, \dots, n-1\})[\mu(x) = x + 1]) \wedge (\mu(n) = 1) \quad (55)$$
$$((\forall i \in \{1, \dots, n-1\})[\pi(i) = i + 1]) \wedge (\pi(n) = 1) \quad (56)$$

Now, let $P$ satisfy

$$(\forall i)[t_1(P_i) = i] \quad (57)$$
$$(\forall i)(\forall m \in \{1, \dots, n-1\})[t_{m+1}(P_i) = \mu(t_m(P_i))] \quad (58)$$

Therefore, $\mu P = \pi P$. Given any $x$, neutrality and anonymity together imply that

$$f(P)(x) = f(\mu P)(\mu(x)) = f(\pi P)(\mu(x)) = f(P)(\mu(x)) \quad (59)$$

Now, by equation 59, $f(P)(1) = \cdots = f(P)(n)$. Hence, $f(P)(x) = 1/n$ for all $x$. By Claim 2.2, every $h$ satisfies $h(P)(x) = 1$ for some $x$. Since $f(P)(x) > 0$ for all $x$, it follows that $\{x \mid (\exists h)[h(P)(x) = 1]\} = A_n$. Thus, $|H| \geqslant n$. Consider any $v$ and $G$. By self-equivalence, there exists some $R$ such that $\text{supp}(f(R)) = H$. Fix any such $R$. By Claim 2.1, $H \subseteq B(R)$. Therefore, $|H| \leqslant |B(R)| \leqslant n$. Hence, $|H| = n$.

CLAIM 2.4. $(\forall h)[\lambda(h) = 1/n]$.

The proof is direct. Consider the profile $P$ constructed in the proof of Claim 2.3. By Claim 2.2, every $h$ satisfies $h(P)(x) = 1$ for some $x$. Since $f(P)(x) = 1/n$ for all $x$, it follows that $\lambda(h) \leqslant 1/n$ for all $h$. By Claim 2.3, $|H| = n$. Thus, $\lambda(h) = 1/n$ for all $h$.

CLAIM 2.5. $f = \varphi$.

The proof is direct. Consider any $\tau$, $P$, and $v$. Let $G = H \cup \mathcal{F}_d$. By equation 29, there exists some $R$ such that $\text{supp}(f(R)) = H$. Fix any such $R$. Given any $i$, let $h_i = t_1(R_i)$ be *voter $i$'s top-ranked voting* *rule at* $R$. By Claim 2.1, $H \subseteq \{h_1, \dots, h_n\}$, so $|H| \leqslant |\{h_1, \dots, h_n\}| \leqslant n$. By Claim 2.3, $|H| = n$, so $H = \{h_1, \dots, h_n\}$ and $h_i \neq h_j$ whenever $i \neq j$. Given any $i$, $h_i R_i^{\nu} d_i$. Therefore,

$$v_i(t_1(P_i)) = u_i(d_i(P)) \leqslant u_i(h_i(P)) \leqslant v_i(t_1(P_i)) \quad (60)$$

Since $v_i(x) < v_i(t_1(P_i))$ for all $x \neq t_1(P_i)$, it follows that $h_i(P) = d_i(P)$. Given any $x$, equation 61 follows from the fact that $\lambda$ induces $f$, equation 62 follows from Claim 2.4 and $H = \{h_1, \dots, h_n\}$ with distinct $h_i$, equation 63 follows from $h_i(P) = d_i(P)$ for all $i$, and equation 64 follows from equation 5:

$$\begin{aligned} f(P)(x) &= \textstyle\sum_{h\in H} \lambda(h)h(P)(x) && (61)\\ &= (1/n)\textstyle\sum_{i\in N} h_i(P)(x) && (62)\\ &= (1/n)\textstyle\sum_{i\in N} d_i(P)(x) && (63)\\ &= \varphi(P)(x) && (64)\end{aligned}$$

Hence, $f = \varphi$. □

REMARK 1. With the exception of neutrality, the axioms in Theorem 1 are logically independent. Given any $\tau$ and $P$, let $E(P) = \{x \mid (\forall y)(\exists i)[x P_i y]\}$ be the *set of all strictly Pareto undominated alternatives at* $P$. Let $d_1 \in \mathcal{F}$ be *voter 1's dictatorship*, let $\psi \in \mathcal{F}$ be the *uniform random anti-dictatorship*, and let $\sigma \in \mathcal{F}$ be the *voting rule that assigns equal probability to all strictly Pareto undominated alternatives*. Formally,

$$(\forall \tau, P)[d_1(P)(t_1(P_1)) = 1] \quad (65)$$

$$(\forall \tau, P, x)[\psi(P)(x) = (1/n)n_\tau(x, P)] \quad (66)$$

$$(\forall \tau, P, x)[(x \in E(P)) \Rightarrow (\sigma(P)(x) = 1/|E(P)|)] \quad (67)$$

Then, the axioms satisfied by $d_1$, $\psi$, and $\sigma$ are depicted in Table 3. The proofs of the claims in Table 3 are left to the reader because their arguments are very similar to those used throughout the proof of Theorem 1. The exception concerning neutrality follows from $\mathcal{F}_s \subseteq \mathcal{F}_n$. Thus, neutrality can be omitted from Theorem 1, but I retain it to explicitly present anonymity, neutrality, and efficiency as normative democratic requirements.

TABLE 3. LOGICAL INDEPENDENCE

| $f$ | $\mathcal{F}_a$ | $\mathcal{F}_e$ | $\mathcal{F}_n$ | $\mathcal{F}_s$ |
|---|---|---|---|---|
| $d_1$ | **No** | Yes | Yes | Yes |
| $\psi$ | Yes | **No** | Yes | Yes |
| $\sigma$ | Yes | Yes | Yes | **No** |

## 8. CONCLUSION

I conclude the paper by discussing the reasons behind the key modeling choices in self-equivalence.

**Neutrality**. Extending the domain of a nonneutral voting rule to linear order profiles on arbitrary finite sets of voting rules requires identifying those voting rules with alternatives through some bijection. Since different bijections may produce different outcomes, defining self-equivalence for nonneutral voting rules requires specifying which bijections are considered, with potential implications for the meaning of the axiom. For neutral voting rules, the choice of bijection is inconsequential. Thus, I define self-equivalence for neutral voting rules so that it is independent of this identification and has a clear interpretation. The components of the inducing lottery and the available rival voting rules need not be neutral, since evaluating them at linear order profiles on alternatives does not require extending their domains.

**Inducing lotteries**. A voting rule may admit several inducing lotteries, yet self-equivalence simply requires the existence of one such lottery whose recovery remains possible across all admissible situations. Unfortunately, requiring recovery of every inducing lottery is incompatible with anonymity, efficiency, and neutrality.[8] Since the support of the inducing lottery determines which sets of voting rules are admissible, fixing the inducing lottery ensures that the lottery to be recovered and the sets of available voting rules do not depend on voters' linear order profiles or utility profiles.

**Utility profiles**. Since different utility profiles representing the same linear order profile on alternatives may induce different weak order profiles on lotteries, requiring the predicate of self-equivalence to hold for some utility profile would make the stability requirement conditional on voters' preference intensities. Hence, recovery is required for every such utility profile to ensure that, given any weak order profile over lotteries constructed via expected utility, the inducing lottery remains recoverable at some compatible linear order profile. While this requirement makes self-equivalence quite strong, it is crucial to the proof of Claim 2.1.

**Available voting rules**. Requiring the predicate of self-equivalence to hold for arbitrary admissible finite sets of voting rules makes the stability requirement robust to whichever finite set of voting rules is available to society. The requirement that the support of the inducing lottery be a subset of the finite set of available voting rules stems from the fact that recovering that inducing lottery is otherwise impossible.

**Compatible profiles**. Requiring the predicate of self-equivalence to hold at every compatible linear order profile on any admissible finite set of voting rules renders self-equivalence impossible to satisfy.[9] However, the existential requirement admits a conservative interpretation: society resolves individual indifferences in favor of preserving its current voting rule through the recovered lottery because transitions are often costly.

**Koray (2000)**. If a voting rule is *deterministic* (i.e., it always chooses a degenerate lottery), self-equivalence requires it to assign probability one to itself. Further, if the available voting rules must also be deterministic, the compatible linear order profiles on any admissible finite set of voting rules depend only on voters' linear order profiles on alternatives. In that case, self-equivalence reduces to a stochastic variant of Koray's (2000) universal self-selectivity that admits nonneutral rival voting rules.

**ACKNOWLEDGMENTS**. *I am profoundly grateful to Christopher Stapenhurst for several conversations that were fundamental to the development of this paper. I am also indebted to Sreoshi Banerjee, Noémi Cabau, Michael Greinecker, Toygar T. Kerman, M. Remzi Sanver, Anastas P. Tenev, and Róbert Somogyi for their useful comments. Furthermore, I thank seminar participants at the Quantitative Social & Management Sciences Research Centre at the Budapest University of Technology & Economics, the Institute of Economics at the Corvinus University of Budapest, the Institute of Economics at the Centre for Economic & Regional Studies, the Department of Economics at the University of Rovira i Virgili, the Economic Theory Group at the University of Liverpool, the Department of Economic Theory & History at the University of Granada, the Department of Economics at the University of Barcelona, the Kyiv School of Economics, the Faculty of Economic Sciences at HSE University, Plaksha University, the Department of Economics at the University of Ioannina, the Group for Research in Applied Economics, and the Faculty of Economics at the European Research University. Finally, I thank participants at the 2025 European Meeting on Game Theory and the 2025 Conference on Economic Design. All errors are my own.* • **ARTIFICIAL INTELLIGENCE**. *I used ChatGPT 6 Astra Ultra, Google Deep Review, and Refine.ink to check the proof and improve the structure, language, and exposition of the manuscript. I reviewed the tools' outputs*

[8]To see this, let $\tau = n$, let $\lambda(\varphi) = 1$, let $G = \{\varphi\} \cup \mathcal{F}_d$, and let $P$ satisfy $t_1(P_i) \neq t_1(P_j)$ for all $i$ and $j \neq i$. Then, given any $\nu$, $d_i R_i^\nu \varphi$ and $\neg(\varphi R_i^\nu d_i)$ for all $i$. Thus, every $R$ satisfies $\varphi(R)(\varphi) = 0 \neq 1 = \lambda(\varphi)$. Further, since this stronger requirement implies self-equivalence, Theorem 1 implies that no voting rule satisfies it together with the other axioms.

[9]To see this, let $\tau = 2$, and let $h \in \text{supp}(\lambda)$. Given any $P$ and $\nu$, let $g \in \mathcal{F} \setminus \text{supp}(\lambda)$ satisfy $g(P) = h(P)$. Let $G = \text{supp}(\lambda) \cup \{g\}$, consider any $R$, and construct $\tilde{R}$ by swapping $h$ and $g$ in every $R_i$. Since $g(P) = h(P)$, $\tilde{R}_i \subseteq R_i^\nu$ for all $i$. By neutrality, recovery at both profiles would imply $0 < \lambda(h) = f(R)(h) = f(\tilde{R})(g) = \lambda(g) = 0$, which is a contradiction.

*and take full responsibility for the submitted manuscript.* • **CONFLICT OF INTEREST**. *None.* • **DATA**. *None.* • **FUNDING**. *None.*